\documentclass[doublecol]{epl2} 

\usepackage{standalone}
\usepackage[utf8]{inputenc}
\usepackage{amsmath}
\usepackage{graphicx}
\usepackage{soul}
\usepackage{color}
\usepackage{comment}
\usepackage{amssymb}
\usepackage{amsmath}
\usepackage{amsthm}
\usepackage{commath}
\usepackage{float}
\usepackage{graphicx}
\usepackage{mathtools}
\usepackage{hyperref}
\usepackage{xcolor}
\usepackage{xargs}                      
\usepackage{arydshln}

\usepackage{subfigure}
\usepackage{scalerel}
\usepackage{fixltx2e} 
\usepackage{url} 

\title{A model for the Mediated Artificial Square Ice phenomenology}

\author{F. Caravelli}
\date{Jan 2020}
\institute{Theoretical Division and Center for Nonlinear Studies,\\
Los Alamos National Laboratory, Los Alamos, New Mexico 87545, USA\\
}


\abstract{
 In the present Letter we discuss the origin of the vertex population inversion observed experimentally in the mediated Artificial Square Ice \cite{modifiers}.
 An interaction modifier is a disc-shaped magnetic nanoisland (a dot) which is placed at the center of a vertex to mediate the interaction between the nearby islands. We show that the inversion is of entropic origin, and can be explained via the renormalization of the vertex configuration energies due to local interaction between the nanoislands and the dot. We show it in a model with mixed Heisenberg and Ising spins (a spin-dot interaction) that as a function of the island size, entropic effects become important. Because of the renormalization of the energies, we observe a level crossing between \textrm{Type I} and \textrm{Type II} vertices which is qualitatively similar to the experimental results. We also discuss possible implications of spin-dot interactions in the eight- and sixteen- vertex models phase diagrams, using both the exact order parameter in the former case and the numerically inferred one in the latter.
}

\begin{document}

\maketitle

\section{Introduction}  
The last years have seen the use of a variety of interacting magnetic nanostructures~\cite{colloq,Wang1,Bader} in different geometries, and the introduction of artificial materials whose behavior is similar to the one of classical spin ice models. Today, because of an  experimental better understanding of these materials, the interest is shifting from reproducing the behavior of known statistical physics models, to novel ones. In particular, various new phenomena have been investigated \cite{reddim,Heyderman,Canals1,Nisoli1,Morgan,Budrikis,Branford,Ryzhkin,Chern2,Branford,Le,Chern3,Gliga,Nisoli4,reddim,Gilbert2,Bhat}, ranging from topological order \cite{Castelnovo1,topor}, memory in materials \cite{Lammert2,GilbertMem}, disordered systems and slow relaxation \cite{Cugliandolo2,Caravelli2}, novel resistive switching \cite{memristors,caravelli}, and embedding logic circuits in the magnetic substrate \cite{logic1,logic2,logic3,logic4,logic5} just to mention a few.
The level of manipulation obtained of each island is remarkable \cite{gartside,WangYL2,WangYL,colloq,Nisoli4}, which now suggests the study of new types of interactions for novel models \cite{Gilbert,Nisoli8,Schanilec}.

\begin{figure}
    \centering
    \includegraphics[scale=1.2]{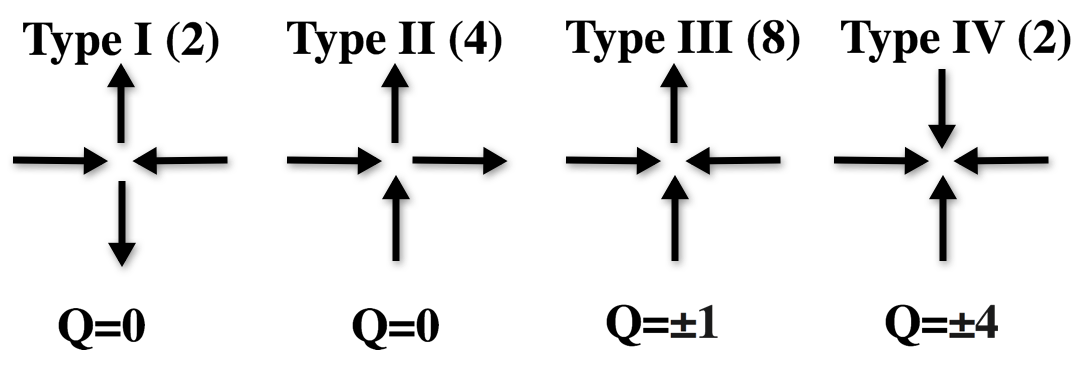}\\
    \caption{Vertex configurations the sixteen vertex model and the associated magnetic charges $Q$, up to a change in the orientation of each spin, for the vertices of \textrm{Type I} - \textrm{Type IV}.}
    \label{fig:vertexenergies}
\end{figure}

A well known example of possible application of artificial spin ice (ASI) is the possibility of having monopole like charges in  spin ices \textit{without} a string tension \cite{Castelnovo2,Mol,Mengotti,Ladak1,Ladak2,Zeissler,Faran1}.
In the case of ASI \cite{Wang1}, however, the key problem is that nanoislands have an asymmetric interaction due to the dipolar nature of the exchange couplings. In the approximation of nearest-neighbor and dipolar-like interactions, the energy of an artificial square ice can be approximated using local energy contributions associated to the interactions between parallel and perpendicular islands $\epsilon_{\perp}>\epsilon_{||}>0$, whose magnitude depends on their distance \cite{NisoliPRL}. 
\begin{figure}
    \centering
    \includegraphics[scale=1.5]{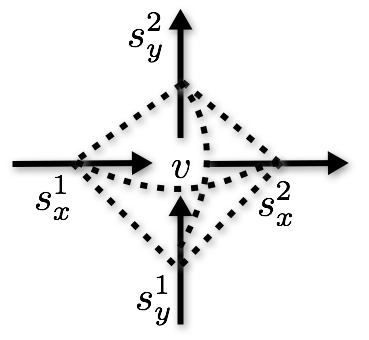}
    \caption{Interactions at a vertex $v$ in an artificial square ice (ASI). There are interaction both between the parallel and perpendicular islands, and characterized by two energies $\epsilon_{||}$ and $\epsilon_{\perp}$.}
    \label{fig:asi}
\end{figure}
Below their Curie temperature $T_c$, each magnetic island $s^i$ acquires an Ising-like spin $s^i=\pm 1$, associated to the internal direction of the magnetization, because of the typical geometrical elongation of the nanoislands. Given this,
let us now briefly explain how the energy of each spin configuration is obtained, considering Fig. \ref{fig:asi}. The artificial square ice is thus described, at the level of the vertex, by the following Hamiltonian based on Ising-like variables, but lying on the plane:
\begin{equation}
    H_{ASI}=-\sum_{v}\Big[ \epsilon_{||}(\sum_{\langle i,j\rangle_v} s^i_{x}s^j_{x}+\sum_{\langle i,j\rangle_v} s^i_{y}s^j_{y} )+ \epsilon_{\perp} \sum_{\langle i,j\rangle_v} s^i_{x} s_{y} ^j\Big]
\end{equation}
where on the plane we have the spin direction to be $\vec s_{x}=s_{-}\hat x$ and $\vec s_{y}=s_{-}\hat y$. Alternatively, one can describe these type of models in terms of the four type of vertex configurations (their energeies) only, as in Fig. \ref{fig:vertexenergies}.

For the square spin ice, it has been noted that vertices have four increasing energies parametrized by $\epsilon_{\perp}$ and $\epsilon_{||}$, with a nomenclature \textrm{Type I},$\cdots$,\textrm{Type IV} respectively. The vertex energies are
    $\epsilon_{I}=-4 \epsilon_{\perp}+2 \epsilon_{||}$,  $\epsilon_{II}=-2 \epsilon_{||}$, \\
    $\epsilon_{III}=0$,
    $\epsilon_{IV}= 4\epsilon_{\perp}+2\epsilon_{||}$,
where $\epsilon_{I}<\epsilon_{II}<\epsilon_{III}<\epsilon_{IV}$. The vertex population in the ground state is determined by this energy hierarchy. We assume that the Boltzmann constant  $\kappa$ is equal to one, and thus measure the energy in the temperature. In these units, we have
$\epsilon_{\perp}\approx0.38675$ and $\epsilon_{||}\approx 0.2735$ for realistic phase diagrams, as noted in \cite{Morrison}.

 Following the discussion above, in order to have monopole-like excitations freely to move in the material, it would be desirable that vertices of \textrm{Type I}and \textrm{Type II}have the same energy.
Recently, ways to overcome this difficulty have been proposed in the literature, using for instance 3-d materials (by raising two of the four islands at a vertex) \cite{ladak} or via intermediate interactions \cite{modifiers}. In the case of the recent publication in \cite{modifiers}, it has been proposed to introduce a ``dot", a disc-shaped islands, at the center of each vertex between four islands, also called interaction modifier. In this paper we discuss this type of mediated interactions.

\begin{figure}[ht!]
    \centering
    \includegraphics[scale=0.45]{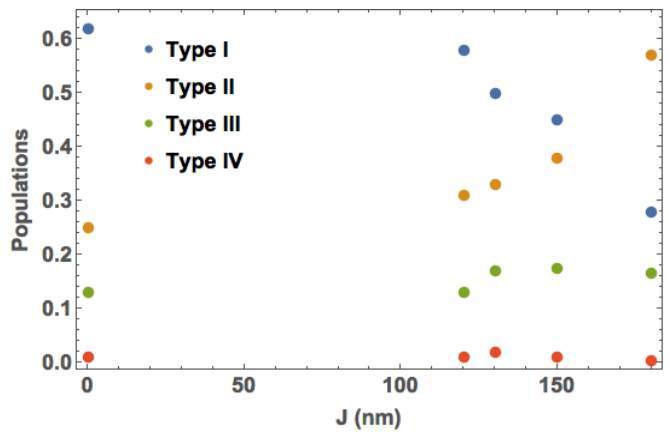}
    \caption{We plot the population inversion between \textrm{Type I} and \textrm{Type II} in the ground state observed experimentally in \cite{modifiers}, as a function of the disc size (in nanometers). It can be seen that for $J\approx 155$ \textrm{Type II} vertices become more present than \textrm{Type I}.}
    \label{fig:paperresults}
\end{figure}

It has been experimentally observed that as the size of the island increases, at low temperature the population of each vertex configuration in the ground state can flip depending the size of the dot \cite{modifiers}, which we report in Fig. \ref{fig:paperresults}. 
As it can be seen, when the size of the dot-island increases, vertices of \textrm{Type I} invert their population with vertices of \textrm{Type II}. It has thus been observed that there is an optimal dot size such that the energy of the two configurations are the same. The purpose of this paper is to introduce a simple model to explain such phenomenology. Such inversion can in principle be explained via a change in the energetics of the vertex configuration due to the presence of the dot.

\section{Spin-Dot as an Ising-Heisenberg interaction}
\begin{figure}
    \centering
    \includegraphics{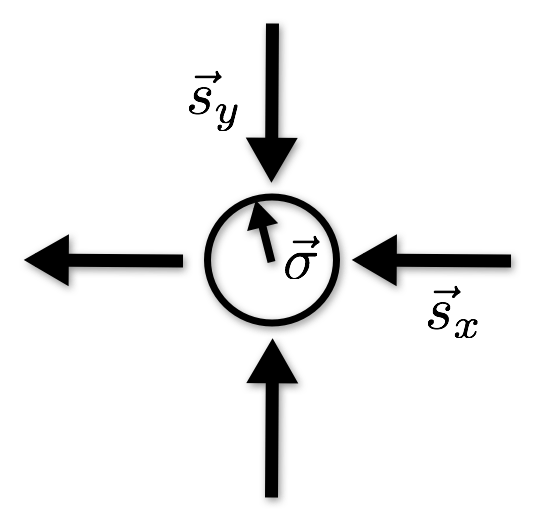}
    \caption{The Ising-Heisenberg type of interaction to the presence of the dot at each vertex. On top of the interaction between the magnetic nanoislands, we also have the interaction between the dot and each island.}
    \label{fig:spindot}
\end{figure}

 As we discuss below,  the presence of the dot at each vertex changes the energy landscape of the model. In particular, we propose to describe this type of interaction with the introduction of extra degrees of freedom at each vertex, in the standard approach to understand Artificial Square Ice via the energetics of the horizontal and vertical islands. First, we note that the dot island lacks any breaking of the horizontal and vertical symmetry. We thus find reasonable assuming that below the Curie temperature such island acquires a two dimensional Heisenberg type spin, as shown in Fig. \ref{fig:spindot}.
Since a disc island sits at each vertex, we suggest to consider the additional interaction between the ASI nanoislands with the dot as mixed spin-Heisenberg interaction of the type
\begin{equation}
    H_{sd}=J \sum_v \left(\sum_{\langle i,j\rangle_v} \vec s^i \cdot \vec \sigma_j \right),
\end{equation}
where $\vec \sigma_j$ is now a two dimensional Heisenberg spin, $\sigma=\cos \theta \hat x+\sin \theta \hat y$.  Here, the coupling energy $J$ effectively can be ascribed to the physical dimension of the dot island. These type of models are known in the literature \cite{Fisher}, and various type of decorations and mappings are possible for arbitrary lattice configurations. However, in the present paper we will see that integrating out the local degrees of freedom will be sufficient.

The total Hamiltonian is thus given by $H=H_{ASI}+H_{sd}$ which we now study. The advantage of using the Hamiltonian above is that the spin-Heisenberg interaction can be exactly integrated out. 
The partition function is thus now generalized by a  
\begin{equation}
    Z=\sum_{\{s_k=\pm 1\}} \int d\theta_i e^{-\beta H_{ASI}-\beta H_{sd}},
\end{equation}
with $\beta$ representing the inverse temperature, $\theta_i$ is the angle of the dot island and $s_x$'s and $s_y$'s are the lattice spin sum.
Since the interactions in $H_{sd}$ are between two different type of spins, the interaction graph is bipartite. Thus, we can integrate out the degrees of freedom of the disc at each vertex, and reabsorb it in the Hamiltonian in terms of the spin configurations. Because of this, the partition function can be written effectively as $Z=  \sum_{\{n_k=0\}}^N \prod_k w_k^{n_k}$, where for $J=0$, $w_i=e^{-\beta \epsilon_i}$, and $N$ is consistent with the constraints. Because of the spin-dot interaction, however the corrected weights can be written as $w_i^c=w_i v_i(\beta)$, where we now study the contribution $v_i$ in detail.

The spin-dot interactions are interesting on their own, as we have a mixing between Ising and Heisenberg spins, and some comments on this matter will be made at the end of this Letter.
First, we note that the integral over each dot island can be performed exactly. In fact, the structure of the interactions  is bipartite, which implies that integrating away the dot island implies an effective interaction between the nano-islands.  We can now write
\begin{eqnarray}
    Z_v^{sd}&=&\int_0^{2\pi} d\theta_v e^{-\beta H_{sd}^v(\theta_v)} \nonumber \\
    &=&\int_0^{2\pi} d\theta_v e^{\beta J(s_{l}+s_{r})\cos \theta+\beta J(s_{b}+s_{u})\sin \theta} \nonumber \\
    &=&2\pi I_0(\beta J \sqrt{4+2s_l s_r+2s_u s_d}) \nonumber \\
    &\equiv& 2 \pi e^{-\log I_0 }= 2\pi e^{-\beta H_{ent}}
\end{eqnarray}
where $I_\nu(x)$ are the Modified Bessel function of the first kind, and we have defined $H_{ent}=\beta^{-1} \log I_0$. We thus see that this term can be re-inserted again inside the energies of the 16-vertex model, but now with an asymmetric energy. The effective correction to the weight of the vertex is thus 
$v_i(\beta)=  I_0(\beta J \sqrt{4+2s_l s_r+2s_u s_d})$, and the energy correction $\delta \epsilon_i=-\beta^{-1} \log v_i(\beta)$ and thus entropic. We aim to show that such simple contribution is enough the explain the vertex inversion observed experimentally. If we include the contributions of the horizontal and vertical interactions,
 we find at finite temperature that the corrections to the energy levels are the form    
\begin{eqnarray}
\epsilon_{I}^\beta&=& \epsilon_{I},    \nonumber \\
\epsilon_{II}^\beta &=& \epsilon_{II}-\beta^{-1} \log I_0(\sqrt{8}\beta J ),\nonumber \\
\epsilon_{III}^\beta &=& \epsilon_{III}-\beta^{-1}\log I_0(2\beta J ),\nonumber \\
\epsilon_{IV}^\beta &=&\epsilon_{IV}.
\end{eqnarray}
We see immediately that in the limit $x\rightarrow 0$, we have $I_0(x)\rightarrow 1$. Since the effective energy is $\delta e\propto \log I_0$, the energy correction of the dot interaction at each vertex is expected not to be important at high temperatures.  This is also observed in the experimental results. We argue that such contribution is entropic, the reason being that the change in the energy configuration is effectively due to the (microscopic) degrees of freedom being integrated out, and that these contribute to the (macroscopic) energy levels of the islands. If such contribution was identical for all vertices, such contribution would not change the energy landscape. Instead, 
it is interesting to note that such effective contribution is asymmetric in the vertex configurations, which has an important effect at low temperatures.
If $J\neq 0$ and sufficiently large, then at low temperatures we get modifications of the energetics for the vertex. For $\beta \gg 1$, we have $I_0(z)\approx \frac{e^z}{\sqrt{2\pi z}}+O\left(\frac{1}{z^{\frac{3}{2}}}\right)$. From this formula we can derive $\delta\epsilon=\beta^{-1} \log I_0(\beta J x)\approx\beta^{-1} \left(\beta J x-\frac{1}{2}\log (\beta J x)\right)$, which can be approximated as $\delta \epsilon\approx J x$ at temperatures close to zero.

Effectively, since the interaction with the dot can be reabsorbed in a change of the energy levels, the entropic change can also be interpreted  as a simple form of renormalization (or rather, screening). If we add these correction to the energy of each vertex, we get low temperature renormalized value for  the vertices configurations of the form
\begin{eqnarray}
    \epsilon_{I}^\infty&=& \epsilon_{I},  \ \ \ \ \   \epsilon_{II}^\infty = \epsilon_{II}-\sqrt{8} J,  \nonumber \\
    \epsilon_{III}^\infty &=& \epsilon_{III}-2 J, \ \ \ \ \    \epsilon_{IV}^\infty =\epsilon_{IV}.
    \label{eq:levels}
\end{eqnarray}

\begin{figure}[ht]
    \centering
    \includegraphics[scale=0.44]{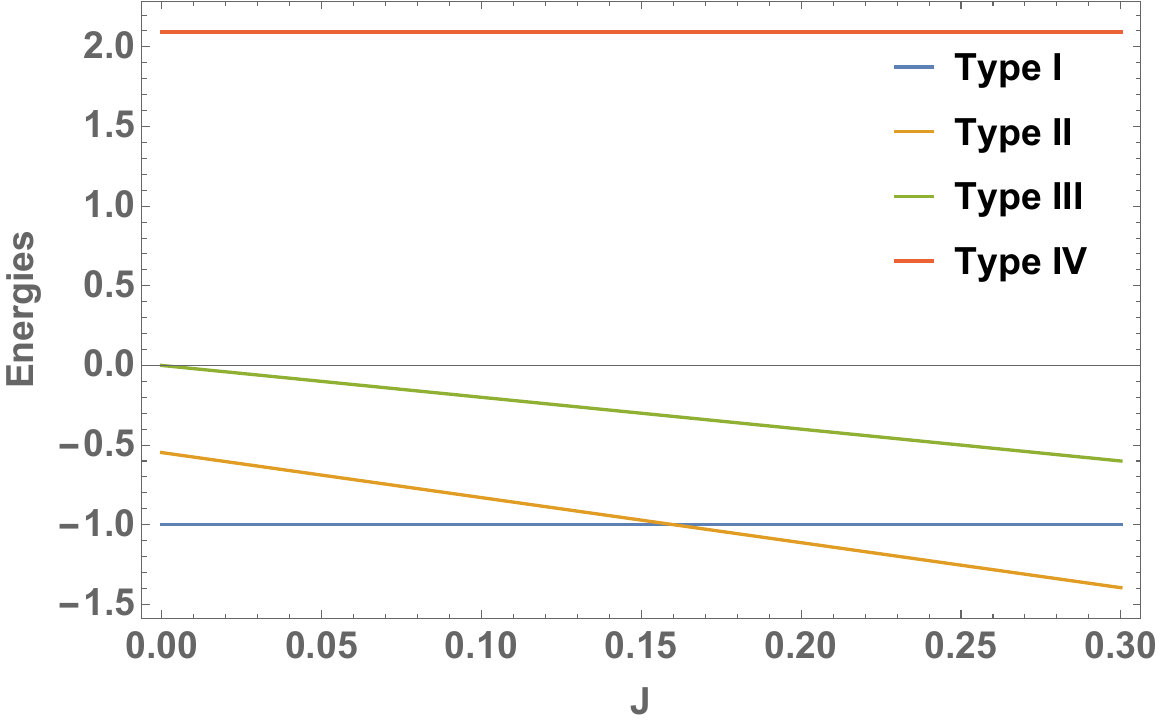}
    \includegraphics[scale=0.435]{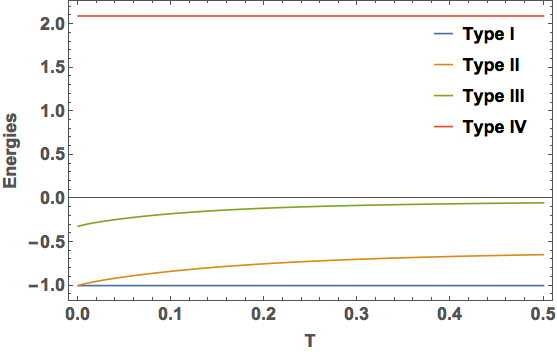}
    \caption{\textit{Top:} Level inversion predicted as an entropic effect due to an asymmetrical change in the energetics of the vertices, for $\beta\rightarrow \infty$ and as a function of $J$. \textit{Bottom:} ``Flow" of the energies with the temperature for $J=J^*$.}
    \label{fig:levelinv}
\end{figure}

Given the formulae in eqns. (\ref{eq:levels}), we can now obtain the main result of this paper; the level inversion can be observed as a function $J$ in Fig. \ref{fig:levelinv} (top). The optimal interaction strength $J^*$ such that $\epsilon_I=\epsilon_{II}$ can immediately be obtained in our model, being
\begin{equation}
    J^*=\frac{\epsilon_{II}-\epsilon_{I}}{\sqrt{8}}.
\end{equation}
Once  we fix $J=J^*$, the effective energies of the vertices as a function of the temperature are easily obtained, and shown in Fig. \ref{fig:levelinv} (bottom). It is important to note that  $J^*$ has been fixed so that $\epsilon_I=\epsilon_{II}$ at $T=0$. However, one could choose $J^*(\beta^*)$ so that the effective energy cross at another finite temperature $T^*>0$. Since the coupling $J$ is connected to the effective size of the island, this is another qualitative agreement with the experimental results of \cite{modifiers}.

\section{Implications for ice models: paramagnetic-ferromagnetic transitions}
\begin{figure}
    \centering
    \includegraphics[scale=0.38]{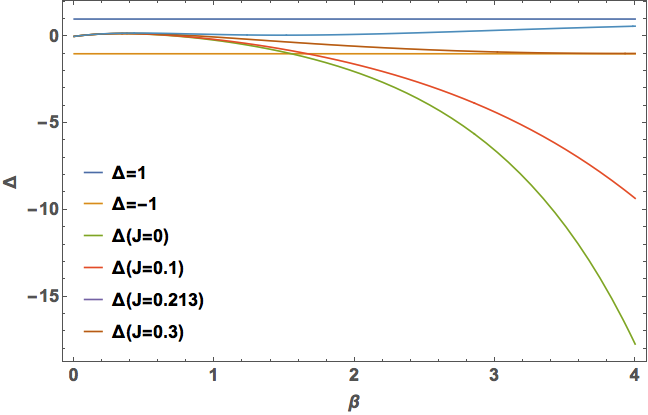}\\
    \includegraphics[scale=0.38]{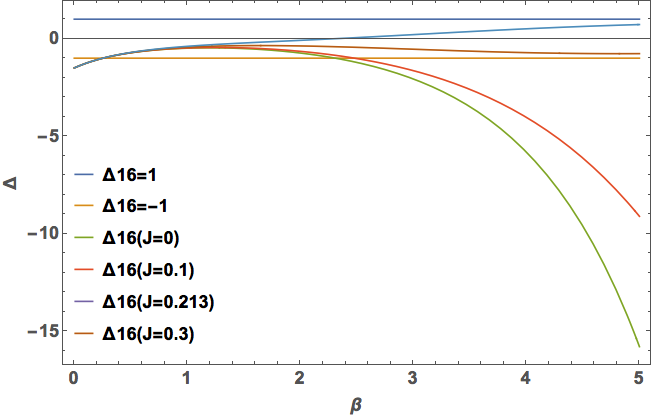}\\
    \caption{ Renormalized $\Delta$ (top) and $\Delta_{16}$ (bottom) for $\epsilon_{\perp}\approx0.38675$ and $\epsilon_{||}\approx 0.2735$, as a function of $\beta$ for various values of $J$ }
    \label{fig:critc}
\end{figure}

Let us now briefly comment on the non-triviality of such interactions for the phase properties of ice models. We first consider a reduction of the model to the case of vertices of only \textrm{Type I}, \textrm{Type II} and \textrm{Type IV}, which is the exactly solvable 8-vertex model \cite{Baxter}. For energies $a=b=\exp(-\beta \epsilon_{II})$, $c=\exp(-\beta \epsilon_{I})$, $d=\exp(-\beta \epsilon_{IV})$, let us define for the case without dot the following order parameter $\Delta=\frac{a^2+b^2-c^2-d^2}{2 (ab+cd)}$.
It is known that if $\Delta>1$, the system is in a ferromagnetic phase, if $\Delta<-1$, the system is in an anti-ferromagnetic phase, and if $-1<\Delta<1$, the system is in a paramagnetic phase. Let us now consider again the choice of the parameters $\epsilon_{\perp}\approx0.38675$ and $\epsilon_{||}\approx 0.2735$.  In Fig. \ref{fig:critc} (top) we plot $\Delta$ for various values of $J$. We can see that for $J=0$, at $T=\infty$ the system is in a paramagnetic phase, and for $T\rightarrow 0$ the system undergoes a transition to the ferromagnetic phase. However, for $J\neq 0$ such picture changes if replace $\epsilon_{k}\rightarrow \epsilon_{k}^\beta$, and for $J>J_c\approx 0.213$ the system remains in a paramagnetic phase. Thus, even for the eight vertex model, the presence of the interaction modifiers affects the phase diagram of the model.\ \ \\

The eight-vertex model is however not a good model to describe the artificial square ice, as all possible vertices are present.
Unfortunately, a similar precise analysis for the sixteen-vertex model cannot be done, as the model is not integrable in its full generality, but only for particular choice of parameters \cite{Wu1,Wu2}. However, numerical investigations and the use of the Bethe-Ansatz suggest that a shifted order parameter $\Delta_{16}$ can be used to understand the critical phases of the sixteen vertex model (with a similar interpretation of $\Delta$), in which $d\rightarrow d+3e$, with $e=\exp(-\beta \epsilon_{III})$ \cite{Levis1,Levis2}, 
$\Delta_{16}=\frac{a^2+b^2-c^2-(d+3e)^2}{2 (ab+c(d+3e))}$. Using the parametrization $\epsilon_{III}=0$ in the case without dot, we have $e=I_0( 2 \beta J)$. The behavior of the parameter $\Delta_{16}$ above, as a function of the inverse temperature $\beta$, is shown in Fig. \ref{fig:critc} (bottom). For $J=0$, the model undergoes a AF-PM-AF series of transitions from low to high temperatures, again for the parameter choices $\epsilon_{\perp}\approx0.38675$ and $\epsilon_{||}\approx 0.2735$. However, for $J>0$ the critical behavior changes, with the system undergoing only one of these transitions (AF-PM) for a critical value of $J_c$. Since for the sixteen vertex model the order parameter $\Delta_{16}$ is not exact, the result above should only be indicative of the importance of the interaction modifiers when all possible vertices are included.

\section{Conclusion.} 
We have shown analytically in a simple model that interaction modifiers can be a powerful mechanism to change the energy landscapes of ice models and provided a way to study these effects in detail using a spin-dot interaction model. The approach we have provided in this paper is, in its simplicity, extremely powerful. In fact, it allows to understand the typical renormalization of the energy of each vertex in artificial spin ice because of the local interaction, and it can help to shape the critical behavior of artificial nanomagnets.Our analysis also suggests that, similar to observed experimentally, there is an optimal size $J^*$ for the dot island such that \textrm{Type I} and \textrm{Type II} are of equal energy. This implies that as a spin is flipped in the ice manifold and two monopoles are generated, these can moved far apart with more spin flips without changing the energy (e.g. there is no string tension). This implies that the energy excitations are effectively only associated to the magnetic charge.

\begin{acknowledgments}
 We would like to thank Cristiano Nisoli for comments on a first draft of this paper, and Michael Saccone for a discussion at MMM Las Vegas. This work was carried out under the auspices of the NNSA of the U.S. DoE at LANL under Contract No. DE-AC52-06NA25396. FC was also financed via DOE-ER grants PRD20170660 and PRD20190195. 
\end{acknowledgments}


\begin{thebibliography}{99}



\bibitem{colloq} C. Nisoli et al.,
Rev. Mod. Phys. 85(1473) (2013)
\bibitem{Wang1} R. F. Wang et al., 
Nature 439(7074):303-6 (2006). 
\bibitem{Bader} S.D. Bader, 
Rev. Mod. Phys., 78(1):1 (2006). 

\bibitem{reddim} I. Gilbert et al., 
Nature Phys. 12, 162-165 (2016)
\bibitem{Heyderman} L. J. Heyderman, R. L. Stamps, 
J. of Phys.: Cond. Mat., 25(36):363201 (2013) 

\bibitem{Canals1} B. Canals et al., 
Nat. Comm. 7 (2016)
\bibitem{Nisoli1} C. Nisoli et al, 
Phys. Rev. Lett., 98(21):217203 (2007)

\bibitem{Morgan} J. P. Morgan et al., 
Nat. Phys. 7(1):75-70 (2010)



\bibitem{Budrikis} Z. Budrikis et al., 
Phys. Rev. Lett 109(30):037203 (2012)
\bibitem{Branford} W. R. Branford et al., 
Science, 335(6076):1597-1600 (2012) 
\bibitem{Ryzhkin}I.A. Ryzhkin. 
Z. Ehks. i Teoret. Fiz., 128(3):559-566 (2005) 



\bibitem{Chern2} G.-W. Chern, P. Mellado, 
EPL 114 (3): 37004 (2016)


\bibitem{Le} B. L. Le et al., 
Phys. Rev. B, 95:060405 (2017)
\bibitem{Chern3} G.-W. Chern, 
Phys. Rev. App. 8(6) : 064006 (2017)
\bibitem{Gliga} S. Gliga et al., 
Phys. Rev. Lett, 110(11):117205 (2013)


\bibitem{Nisoli4} C. Nisoli et al.
Nature Phys.13(3):200-203 (2017)






 \bibitem{Gilbert2} I. Gilbert et al., 
 Nat Phys. 10(9):670-675 (2014)
\bibitem{Bhat} V. S. Bhat et al., 
Phys. Rev. Lett. 111(7):077201 (2013)
\bibitem{Castelnovo1} C. Castelnovo et al.,
Ann. Rev. Condens. Matter Phys., 3(1): 35-55 (2012)
\bibitem{topor} Y. Lao et al., 
Nature Phys. 14, 723-727 (2018)
\bibitem{GilbertMem}  I. Gilbert et al., 
Phys. Rev. B, 92(10):104417 (2015) 
\bibitem{Lammert2} P. E. Lammert et al., 
Nat. Phys., 6(10):786-789 (2010)
\bibitem{Cugliandolo2} D. Levis et al., 
Phys. Rev. Lett., 110(20):207206 (2013)
\bibitem{Caravelli2} F. Caravelli, F. Markopoulou, Phys. Rev. D 86 (2), 024019, 2012
\bibitem{memristors} F. Caravelli et al.,
arXiv:1908.08073  
\bibitem{caravelli} F. Caravelli, J. Carbajal, 
Technologies 6(4):118 (2019)

\bibitem{logic1} G. Csaba et al.,
IEEE Trans. on Nano. 99(4):2009 (2003)
\bibitem{logic2} H. Arava et al., 
Nanotechnology 29, no. 26 265205  (2018)
\bibitem{logic3} J. H. Hensen et al.,
Proc. of ALIFE 2018, pp. 15-22, MIT Press, 10.1162/isal-a-00011 (2018)
\bibitem{logic4} M. T. Niemier et al.,
J. of Phys.: Condensed Matter, 23(49):493202 (2011)
\bibitem{logic5} F. Caravelli, C. Nisoli, arXiv:1810.09190 

\bibitem{gartside} J. C. Gartside et al., 
Nature Nano., 13(1):53-58 (2018)
\bibitem{WangYL2} Y.-L. Wang et al., 
Science 352,  6288: 962-966  (2016)
\bibitem{WangYL} Y.-L. Wang  et al., 
Nature Nano. 13(7): 560  (2018)
\bibitem{Gilbert} I. Gilbert et al.,
Physics Today, 69(7):54-59 (2016)
\bibitem{Nisoli8} C. Nisoli, 
Nature Nano. 13(1):5 (2018)
\bibitem{Schanilec}
V. Schanilec et al., 
arxiv:1902.00452
\bibitem{Castelnovo2}
C. Castelnovo et al.,
Nature  451,  pp 42-45 (2008)

\bibitem{Mol}
 L. A. S. M\'{o}l et al., 
 J. Appl. Phys. 106:063913 (2009).



\bibitem{Mengotti}  E. Mengotti et al., 
Nat. Phys., 7(1):68-74 (2010)
\bibitem{Ladak1} S. Ladak et al., 
Nat. Phys., 6:359-363 (2010)
\bibitem{Ladak2} S. Ladak et al., 
N. J. Phys. 13(2):023023 (2011)
\bibitem{Zeissler} K. Zeissler et al.,
Scient. Rep. 3:1252 (2013)
\bibitem{Faran1}
A. Farhan1 et al., Science Advances 5:2   (2019)
\bibitem{NisoliPRL} C. Nisoli et al., Phys. Rev. Lett. 105 (1):047205 (2010)

\bibitem{Morrison} M J Morrison, et al., 
New J. of Phys., 15(4):045009 (2013)
\bibitem{ladak}
A. May   et al.,
Comm. Physics , 2:13 (2019)
\bibitem{modifiers}
E. \"{O}stman et al.,
Nat. Phys. 14,pp 375-379 (2018)
\bibitem{Fisher} M. Fisher, 
Phys. Rev. 113:4, pp 969-981 (1959)










\bibitem{Baxter}  R. Baxter, Exactly Solved Models in Statistical Mechanics, Academic Press (London), 1989
\bibitem{Wu1}
F. Y. Wu,
Phys. Rev. Lett. 24, 1476 (1970); Erratum Phys. Rev. Lett. 25, 902 (1970)
\bibitem{Wu2} F. Y. Wu, 
Phys. Rev. B 6:5, (1972)

\bibitem{Levis1} D. Levis, 
PhD thesis,
Universite Pierre et Marie Curie, Paris, France (2012)
\bibitem{Levis2} D. Levis et al.,
Phys. Rev. Lett. 110, 207206 (2013)
\end{thebibliography}

\end{document}